\documentclass[12pt,twoside,bezier]{article}

\font\eightrm=cmr8
\def\beq{\begin{equation}}
\def\eeq{\end{equation}}
\begin{document}


\begingroup
\def\thirteen{\large\bf}
\def\ten{\bf}
\def\eight{\footnotesize}
\baselineskip 15pt
\thispagestyle{empty}
\baselineskip 15pt

\centerline{\ten Centre de Physique
Th\'eorique\footnote{\eight Unit\'e Propre de Recherche
7061}, CNRS Luminy, Case 907}

\centerline{\ten F-13288 Marseille -- Cedex 9}

\vskip 2truecm

\centerline{\thirteen FACET SHAPES IN A WULFF CRYSTAL}

\bigskip

\centerline{{\bf 
Salvador MIRACLE-SOLE\footnote{\eight 
Centre de Physique Th\'eorique, CNRS, Marseille}
}}

\vskip 2truecm

\centerline{\bf Abstract} 

\medskip

According to the Wulff construction the shape of
the equilibrium crystal is determined by the surface tension
considered as a function of the interface orientation.
We present some (conjectured) approximate solutions 
and some rigorous results concerning this function, 
in the case of a lattice gas,
and apply them to study the shape of the equilibrium crystal 
and, in particular, the shape of the facets of this crystal.


\bigskip

Published in {\it Mathematical Results in Statistical Mechanics},
editors
S. Miracle-Sole, J. Ruiz, V. Zagrebnov,  
World Scientific, Singapore, 1999 
(ISBN 981-02-3863-0), pp.\ 83--101. 
Proceedings of the Satellite Colloquium of STATPHYS 20 
held in 
Marseille, France, July 27--31, 1998. 

\bigskip



\bigskip

\noindent Number of figures: 4

\bigskip


\noindent CPT-98/P. 3741

\bigskip



\endgroup
\newpage

\section{Introduction}

\def\tt{\bf}
\def\supp{\mathop{\rm supp}}
\def\dist{\mathop{\rm dist}}

In a first approximation one can model the interatomic 
forces in a crystal by a lattice gas. 
In a typical two-phase equilibrium state there is, 
in these systems, a 
dense component, which can be identified as the crystal 
phase, and a dilute component, which can be identified as the 
vapor phase. 
The underlying lattice structure implies that the 
crystal phase is anisotropic, while this assumption, 
though unrealistic for the vapor phase, should be 
immaterial for the description of the crystal-vapor 
interface \cite{rottman}, \cite{bejeren}. 
As an illustrative example of such systems, 
the ferromagnetic Ising model will be considered.

The Ising model is defined on the $d$-dimensional 
cubic lattice ${\cal L}={\bf Z}^d$, 
with configuration space $\Omega = \{-1,1\}^{\cal L}$. 
The value $\sigma(i)=\pm1$ is the spin at the site $i\in{\cal L}$.  
The occupation numbers  
$n(i)=(1/2)(\sigma(i)+1)$ $=0$ or $1$ 
give the lattice gas version of this model.
The energy of a configuration
$\sigma_{\Lambda} = \{\sigma(i),i\in\Lambda\}$, 
in a finite box $\Lambda\subset{\cal L}$,
under the boundary conditions ${\bar\sigma}\in\Omega$, 
is
\beq
H_{\Lambda}(\sigma_{\Lambda}\mid{\bar\sigma})
= - J \sum_{\langle i,j\rangle\cap\Lambda\not=\emptyset}
\sigma(i)\sigma(j) 
\label{A1}\eeq
where 
$\langle i,j \rangle$ are pairs of nearest neighbor sites 
and $\sigma(i) = {\bar\sigma}(i)$ if 
$i\not\in\Lambda$.
The partition function, at the inverse temperature 
$\beta=1/kT$, is given by
\beq
Z^{\bar\sigma}(\Lambda)
=\sum_{\sigma_{\Lambda}}\exp \big(-\beta
H_{\Lambda}(\sigma_{\Lambda}\mid{\bar\sigma})\big)
\label{A2}\eeq
The limit  
\beq
f(\beta)=\lim_{\Lambda\to\infty}- {1\over{\beta
\vert\Lambda\vert}}\ln\, Z^{\bar\sigma}(\Lambda)
\label{A3}\eeq
is independent of the boundary conditions
and defines the free energy per unit volume. 

This model presents, at low temperatures
$T<T_c$, where $T_c$ is the critical temperature, 
two distinct thermodynamic pure phases.
This means two extremal translation invariant Gibbs states, 
which correspond to the limits, when $\Lambda\to\infty$,
of the finite volume Gibbs measures
\beq
Z^{\bar{\sigma}}(\Lambda) ^{-1} \exp
\big( -\beta H_{\Lambda}(\sigma_{\Lambda}\mid\bar{\sigma})\big)
\label{A4}\eeq 
with boundary conditions ${\bar\sigma}$ respectively 
equal to the ground configurations $(+)$ and $(-)$ 
(respectively, ${\bar\sigma}(i) = 1$ 
and ${\bar\sigma}(i) = -1$, for all $i\in{\cal L}$). 
On the other side, if $T\ge T_c$,  
the Gibbs state is unique.

Each configuration inside $\Lambda$ can be described 
in a geometric way by specifying the set of Peierls contours 
which indicate the boundaries between the regions of 
spin $1$ and the regions of spin $-1$.   
Unit square surfaces are placed 
midway between the pairs of nearest-neighbor sites $i$ 
and $j$, perpendicularly to these bonds, whenever 
$\sigma(i)\sigma(j)=-1$. 
The connected components of this set are 
the Peierls contours. 
Under the boundary conditions $(+)$ and $(-)$, 
the contours form a set of closed surfaces. 
They can be viewed as defects, or excitations,
with respect to the ground states of the system 
(the constant configurations $1$ and $-1$), 
and are a basic tool for the investigation
of the model at low temperatures.

In order to study the interface between the two pure phases
one needs to construct a state describing the coexistence
of these phases.
Let $\Lambda$ be a parallelepiped of sides 
$L_1,L_2,L_3$, parallel to the axes,
and centered at the origin of ${\cal L}$, 
and let ${\bf n}=(n_1,n_2,n_3)$ 
be a unit vector in ${\bf R}^3$, such that $n_3\ne 0$. 
Introduce the mixed boundary conditions $(\pm,{\bf n})$, 
for which 
\beq
{\bar\sigma}(i) = 
\cases{
1  &if\quad $i\cdot{\bf n}\geq 0$ \cr
-1 &if\quad $i\cdot{\bf n}<0$ \cr}  
\label{A5}\eeq 
These boundary conditions force the system to produce a
defect going trans\-versally through the box $\Lambda$,
a big Peierls contour that can be interpreted as the  
microscopic interface.
The other defects that appear above and below the 
interface can be described by closed contours
inside the pure phases.

The free energy, per unit area, due to the presence of the 
interface, is the surface tension. 
It is defined by
\beq
\tau({\bf n})=
\lim_{L_1,L_2\to\infty}\, \lim_{L_3
\to\infty} \, -{{n_d}\over{\beta L_1L_2}}
\ln\, {Z^{(\pm,{\bf n})}(\Lambda)\over 
Z^{(+)}(\Lambda)} 
\label{A6}\eeq 
In this expression the volume contributions 
proportional to the free energy of the coexisting phases, 
as well as the boundary effects, cancel, and only 
the contributions to the free energy due to the interface 
are left. 

\medskip

{\tt Theorem 1. } {\it 
The thermodynamic limit $\tau ({\bf n})$, 
of the interfacial free energy per unit area, exists, 
and is a non negative bounded function of ${\bf n}$. 
Its extension by positive homogeneity,  
$f({\bf x})=\vert{\bf x}\vert\, \tau({\bf x}/ \vert 
{\bf x}\vert)$ 
is a convex function on ${\bf R}^3$.}

\medskip

A proof of these statements has been given in ref.\ \cite{MMR} 
using correlation inequalities 
(this being the reason for their validity for all $\beta$) and,
in fact, the Theorem holds 
for a large class of lattice systems.
The convexity of $f$ is equivalent to the fact that
the surface tension $\tau$ satisfies a thermodynamic 
stability condition called
the pyramidal inequality (see \cite{MMR}, \cite{DSa}).

Moreover, for the Ising model we know, 
from Bricmont et al.\ \cite{BLP}, Lebowitz and Pfister 
\cite{LP}, and the convexity condition, 
that $\tau ({\bf n})$ is strictly positive for
$T<T_c$ and that it vanishes if $T\ge T_c$.

The shape of an equilibrium crystal is obtained,
according to the Gibbs thermodynamic principle,
by minimizing the total surface free energy
associated to the crystal-medium interface.
The solution to this problem, known under
the name of Wulff construction, is the following set
\beq
{\cal W}=\{{\bf x}\in{\bf R}^3 : 
{\bf x}\cdot{\bf n}\le\tau({\bf n})\hbox{ for every } 
{\bf n}\in{\bf S}^{2}\}
\label{A7}\eeq
Notice that the problem is scale invariant, so that if we solve
it for a given volume of the crystal phase, we get the 
solution for other volumes by an appropriate scaling.
The set $\cal W$, which will be called the 
Wulff shape, gives the optimal shape for the crystal.

The Wulff construction can also be viewed as a 
geometric version of the Legendre transformation.
Consider the function $f({\bf x})$ defined  
in Theorem 1.
From definition (\ref{A7}), we get
\beq
{\cal W} =\{{\bf x}\in{\bf R}^d : f^*({\bf x})\le0\}
\label{A8}\eeq 
where $f^*$ is the Legendre transform
\beq
f^*({\bf x}) = \sup_{\bf y}\big({\bf x}\cdot{\bf y}
-f({\bf y})\big)
\label{A9}\eeq
Actually $f^*({\bf x})=0$, if ${\bf x}\in{\cal W}$,
and $f^*({\bf x})=\infty$, otherwise.

We next introduce a function $\varphi$ on ${\bf R}^2$ 
such that the graph of
$x_3= \varphi(x_1,x_2)$, for $x_3>0$,
coincides with the boundary  $\partial{\cal W}$
of the crystal shape.
Since ${\cal W}$ is a convex body, 
symmetric with respect to the origin,
$\varphi$ is a concave function, and
\beq 
{\cal W}=\{{\bf x}\in{\bf R}^3 : 
-\varphi(-x_1,-x_2) \le x_3 \leq \varphi (x_1,x_2)\} 
\label{A10}\eeq 
This means that $-\varphi$
is the Legendre transform of the
projected surface tension $\tau_p=(1/n_3)\tau$, 
considered as a function on ${\bf R}^2$
of the slopes $u_1=n_1/n_3,u_2=n_2/n_3$.  
In other words,
\beq 
\tau_p (u_1,u_2) = f(u_1,u_2,1) 
\label{A11}\eeq 
Indeed, from equations (\ref{A8}) and (\ref{A9}), we see that 
\beq 
-\varphi(u_1,u_2) =
\sup_{u_1,u_2}\big(x_1u_1+x_2u_2-\tau_p(u_1,u_2)\big) 
\label{A12}\eeq

Formula (\ref{A12}) is known as the Andreev construction
\cite{Andreev}).
The interest of this approach comes from the fact
that $\varphi$, and hence, 
the crystal shape itself, may be regarded as the free 
energy associated to a certain statistical mechanical 
Gibbs ensemble. 
We shall consider this ensemble in sections 2 and 3.

\section{The two dimensional model}

We next consider the Ising model on a square lattice,
with two interaction parameters, 
$J_1$ in the horizontal direction and
$J_2$ in the vertical direction.
There is in this model an exact expression
for the surface tension $\tau({\bf n})$.
However, we shall first discuss an approximate expression,
which may represent this quantity in the low temperature region.
It is obtained by restricting the sums in formula (\ref{A6}) to the
ground configurations.

Let ${\cal G}^{\bar\sigma}(\Lambda)$ be the set of 
ground configurations in $\Lambda$ under the boundary 
conditions ${\bar\sigma}$, and let ${\cal N}^{\bar\sigma}(\Lambda)$
and $E^{\bar\sigma}_0(\Lambda)=
H_{\Lambda}(\sigma\vert(\pm,{\bf n}))-H_{\Lambda}(+\vert(+))$
be the number and the relative energy of such configurations.
The set ${\cal G}^{+}(\Lambda)$ contains only the
configuration $(+)$ and, therefore, 
\begin{eqnarray}
\tau({\bf n}) &=&
\lim_{L_1\to\infty}\lim_{L_2\to\infty}-{{n_2}\over{\beta L_1}}
\ln\sum_{\sigma\in{\cal G}^{(\pm,{\bf n})}(\Lambda)}
e^{-\beta H_{\Lambda}(\sigma\vert(\pm,{\bf n}))+ 
\beta H_{\Lambda}(+\vert(+))} \nonumber\\ 
&=& \lim_{L_1\to\infty}\lim_{L_2\to\infty}\,-{{n_2}\over{L_1}}
\big(E^{(\pm,{\bf n})}_0(\Lambda)-\beta^{-1}\ln
{\cal N}^{(\pm,{\bf n})}(\Lambda)\big) \label{A13}
\end{eqnarray}

Notice that, in dimension two, if 
${\bf n}=(-\sin\theta,\cos\theta)$
and $\tau(\theta)=\tau({\bf n})$, then $v=-\tan\theta$ and the
projected surface tension is $\tau_p(v)=\tau(\theta)/\cos\theta$.
The configurations in ${\cal G}^{(\pm,{\bf n})}(\Lambda)$
contain only one Peierls contour, the microscopic interface,
a polygonal line joining two fixed points in the boundary of
$\Lambda$.
This line is cut only once by all straight lines parallel to the
diagonal $i_1-i_2=0$. 
It can then be described by $N$ integers
\beq
\phi(0),\phi(1),\dots,\phi(N)
\label{A14}\eeq
such that
\beq
n(i)=\phi(i)-\phi(i-1)=\pm 1,\quad i=1,\dots,N
\label{15}\eeq
specifying the heights, in units $1/\sqrt{2}$, over the
line $i_1+i_2=0$, of the
extremities of the $N$ consecutive unit segments which
compose the polygonal line (see Figure 1). 
Writing $u=\tan(\pi/2+\theta)$,
the boundary conditions are 
\beq
\phi(0)=0,\quad\phi(0)-\phi(N)=uN
\label{A16}\eeq


\bigskip\bigskip

\setlength{\unitlength}{1mm}

\begin{center}
\begin{picture}(60,60)

\put(5,0){\line(0,1){5}}\put(10,0){\line(0,1){10}}
\put(15,0){\line(0,1){15}}\put(20,0){\line(0,1){10}}
\put(25,0){\line(0,1){5}}\put(30,0){\line(0,1){10}}
\put(35,0){\line(0,1){15}}\put(40,0){\line(0,1){20}}
\put(45,0){\line(0,1){25}}\put(50,0){\line(0,1){20}}
\put(55,0){\line(0,1){25}}\put(60,0){\line(0,1){63}}
\put(15,15){\line(1,1){45}}\put(40,40){\line(1,-1){15}}
\put(40,40){\line(1,0){20}}
\thicklines
\put(0,0){\line(3,1){60}}\put(0,0){\line(1,0){63}}
\put(0,0){\line(1,1){15}}\put(15,15){\line(1,-1){10}}
\put(25,5){\line(1,1){20}}\put(45,25){\line(1,-1){5}}
\put(50,20){\line(1,1){5}}\put(55,25){\line(1,-1){5}}
\put(0,0.2){\line(1,1){15}}\put(15,15.2){\line(1,-1){10}}
\put(25,5.2){\line(1,1){20}}\put(45,25.2){\line(1,-1){5}}
\put(50,20.2){\line(1,1){5}}\put(55,25.2){\line(1,-1){5}}
\put(63,19){$uN$}
\put(63,39){${{1+u}\over2}N$}\put(63,59){$N$}
\put(-1,-5){$0$}\put(58,-5){$N$}
\end{picture}
\end{center}

\bigskip\bigskip

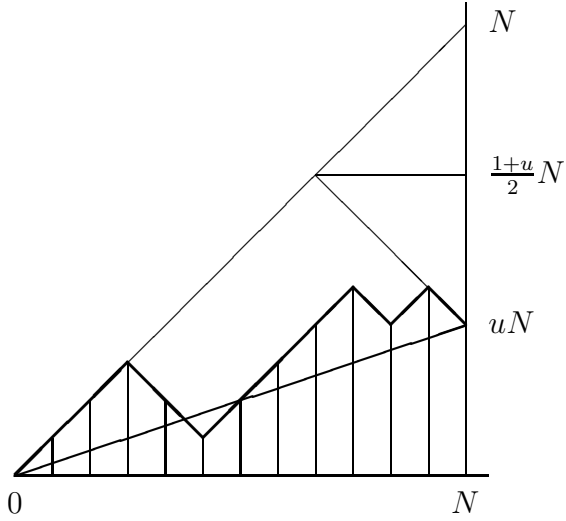
\centerline{\eightrm Figure 1. Description of
a ground microscopic interface in 2 dimensions.}

\bigskip\bigskip


Under these boundary conditions we have that $n(i)=1$,
${{1+u}\over2}N$ times, and $n(i)=-1$, ${{1-u}\over2}N$ times.
All these configurations have the same energy
\beq
E^u(N)=
2J_1{\scriptstyle{1+u}\over2}N+2J_1{\scriptstyle{1-u}\over2}N
\label{A17}\eeq
and their number is 
\beq
{\cal N}^u(N)=\Big({N\atop{{{1+u}\over2}N}}\Big)
\label{A18}\eeq
Therefore, from equation (\ref{A13}), we get 
\begin{eqnarray}
\tau_p(u) &=& J_1({\scriptstyle1+u})+J_2({\scriptstyle1-u})-
\lim_{N\to\infty}{1\over{\beta N}}\ln{{N!}\over{({{1+u}\over2}N})!
({{1-u}\over2}N)!} \nonumber\\
&=& J_1({\scriptstyle1+u})+J_2({\scriptstyle1-u})+
{1\over\beta}\Big({\scriptstyle{1+u}\over2}
\ln{\scriptstyle{1+u}\over2}+
{\scriptstyle{1-u}\over2}
\ln{\scriptstyle{1-u}\over2}\Big)
\end{eqnarray}

In order to obtain the equilibrium crystal shape
we use the Andreev construction (\ref{A12}).
In dimension two, we have
\beq
-\varphi(\xi)=\sup_u(\xi u-\tau_p(u))
\label{A20}\eeq
From $u_0$, solution of
\beq
\xi ={{d\tau_p}\over{du}}=J_1-J_2+{1\over{2\beta}}
\ln{{\scriptstyle1+u}\over{\scriptstyle1-u}}
\label{A21}\eeq
that is
\begin{eqnarray}
{{\scriptstyle1+u_0}\over{\scriptstyle1-u_0}}
&=&e^{2\beta \xi +2\beta J_1-2\beta J_2} \\
u_0 &=&{{e^{2\beta \xi +2\beta J_1-2\beta J_2}-1}\over
{e^{2\beta \xi +2\beta J_1-2\beta J_2}+1}} 
\end{eqnarray}
we find
\begin{eqnarray}
\varphi(\xi ) &=& -\xi u_0+\tau_p(u_0) \nonumber\\
&=& -\xi u_0+J_1+J_2+u_0(J_1-J_2)+{1\over{2\beta}}
\Big(\ln{{\scriptstyle1+u_0}\over{\scriptstyle1-u_0}}+
2\ln{{\scriptstyle1-u_0}\over2}
+u_0\ln{{\scriptstyle1+u_0}\over{\scriptstyle1-u_0}}\Big)\nonumber\\
&=& -\xi u_0+J_1+J_2+u_0(J_1-J_2)+(1+u_0)(\xi +J_1-J_2)
+{1\over{\beta}}\ln{{\scriptstyle1-u_0}\over2} \nonumber\\
&=& \xi+2J_1-{1\over{\beta}}\ln\big(1+
e^{2\beta\xi+2\beta J_1-2\beta J_2}\big)
\end{eqnarray}

Next, we define the partition function 
\beq
\Xi_N(\xi)=
\sum_{\phi}e^{-\beta H(\phi)+\beta\xi(\phi(N)-\phi(0))}
\label{A25}\eeq
where the sum runs over all configurations with $\phi(0)=0$.
We can interpret (\ref{A25}) as a ``grand canonical'' 
partition function 
with respect to the interface boundaries, and the restriction
(\ref{A16}) in (\ref{A13}) as a ``canonical'' constraint.
Since 
\begin{eqnarray}
\phi(N)-\phi(0)&=&\sum_{i=1}^N n(i) \label{A26}\\
H(\phi)&=&2J_1\delta(n(i)-1)+2J_2\delta(n(i)+1) \label{A28}
\end{eqnarray}
we have
\beq
\Xi_N(\xi) 
=\big(e^{-\beta(2J_1-\xi)}+e^{-\beta(2J_2+\xi)}\big)^N
\label{A29}\eeq
and the corresponding free energy is
\begin{eqnarray}
\varphi(\xi)&=&\lim_{N\to\infty}
-{1\over{\beta N}}\ln \Xi_N(\xi) \label{A30}\\
&=&-{1\over{\beta}}\ln 
\big(e^{-\beta(2J_1-\xi)}+e^{-\beta(2J_2+\xi)}\big) \label{A31}\\
&=&\xi+2J_1-{1\over{\beta}}\ln 
\big(1+e^{2\beta\xi+2\beta J_1-2\beta J_2}\big) \label{A32}
\end{eqnarray}

Expressions (\ref{A32}) and (\ref{A25}) coincide.
This means that $\varphi$, the free energy defined by (\ref{A30}),
is the Legendre transform of $\tau_p$, in agreement
with the equivalence of the corresponding ``canonical'' and
``grand canonical'' ensembles. 
This fact illustrates the remark at the end of the introduction.

Putting $\eta=\eta(\xi)=\varphi(\xi)$, equation (\ref{A32}) 
can be written 
\beq
e^{\beta(\eta-\xi-2J_1)}+e^{\beta(\eta+\xi-2J_2)}=1
\label{A33}\eeq
for $\eta>0$.
Then, by
using the initial coordinates $x=\eta+\xi$, $y=\eta-\xi$,
and the symmetries of the figure, we obtain the
following expression for the Wulff shape
\beq
\exp(\beta\vert x\vert-2K_1)+\exp(\beta\vert y\vert-2K_2)\le1 
\label{A34}\eeq
Here $K_i=\beta J_i$, for $i=1,2$. 
This is the approximate solution.
We next describe the exact solution. 

It has been shown that
\beq
y=y(x)=\sup_v(xv-\tau_p(v))
\label{A35}\eeq
is, up to a factor, the Onsager function ${\hat\gamma}(\omega)$ 
for an imaginary argument, namely, 
$y(x)=\beta^{-1}{\hat\gamma}(-i\beta x)$.
This means that
\beq
\cosh\beta y=\cosh2K_1\cosh2K^*_2-\sinh2K_1\sinh2K^*_2\cosh\beta x
\label{A36}\eeq
where
\beq
K^*_i=-({1/2})\ln\tanh K_i
\label{A37}\eeq
is the dual interaction constant ($i=1,2$).
This is a result due to Abraham (see \cite{abraham}).
Taking into account that relation (\ref{A37}) implies
$\cosh2K_i^*$ $\tanh2K_i=1$ and $\sinh2K_i^*\sinh2K_i=1$,
we get
\beq
{{\tanh2K_2}\over{\cosh2K_1}}\cosh\beta x +
{{\tanh2K_1}\over{\cosh2K_2}}\cosh\beta y\le1
\label{A38}\eeq
as the exact expression for the Wulff shape. 
This simple expression has not appear in the literature,
as far as we know.
Several authors (see, for instance, \cite{avron})
have used the more complicated formula for the surface
tension $\tau({\bf n})$, which includes implicit functions, 
instead of the function $y(x)$.

The critical inverse temperature $\beta_c$ satisfies
$K_1=K_2^*$ or $K_2=K_1^*$ or, equivalently, 
\beq
\sinh2K_1\sinh2K_2=1
\label{A39}\eeq
In the symmetric case, $K_1=K_2=K$, we have $\sinh2K_c=1$
and $K_c=(1/2)\ln(\sqrt{2}+1)=0.440687$.
Now, notice that the left hand side of (\ref{A38}) is
\beq
\ge{{\tanh2K_2}\over{\cosh2K_1}}+
{{\tanh2K_1}\over{\cosh2K_2}}
={{\sinh2K_2 + \sinh2K_1}\over{\cosh2K_1\cosh2K_2}}
\label{A40}\eeq
The value of expression (\ref{A40}) equals $1$ for $\beta=\beta_c$, 
because
\begin{eqnarray}
(\sinh2K_2+\sinh2K_1)^2= 2+\sinh^2 2K_1+\sinh^2 2K_1
\hphantom{000000} \nonumber\\
=(1+\sinh^2 2K_1)(1+\sinh^2 2K_1)=
\cosh^2 2K_1\cosh^2 2K_2
\end{eqnarray}
This implies that the Wulff shape reduces to the empty set
for $\beta\le\beta_c$. It has a finite positive volume for
$\beta\ge\beta_c$.

A comparison of expressions (\ref{A33}) and (\ref{A38}) 
shows that the error
made when considering the approximate solution (\ref{A33}) decreases
exponentially in $\beta$ when $\beta\to\infty$.
However, the shape given by (\ref{A33})
presents four corners (when the boundary cuts the axes),
while that corresponding to (\ref{A38}) has a smooth boundary.
The same bound on the error can be proven by means of a low
temperature cluster expansion for the function $y$, 
independently of the exact solution.
This error is less than 1 \% for temperatures less than $(1/2)T_c$.
In the symmetric case the shape defined by (\ref{A33}) becomes the
empty set at $K'_c=(1/2)\ln2=0.346574$.
We have $K'_c < K_c$.

\section{
The three dimensional model:\hfill\break 
Approximate solution}

In this section we consider the Ising model on a cubic lattice
and study the problem of finding in this case 
an approximate solution for the equilibrium crystal shape, 
analogous to that of section 2.
The system so obtained is very interesting in itself, 
though its validity as a low temperature approximation 
is not known.

The ground configurations in ${\cal G}^{(\pm,{\bf n})}(\Lambda)$
contain only one Peierls contour, the microscopic interface.
This contour is a surface made of unit squares such that its
boundary is a fixed line determined by the boundary conditions 
$(\pm,{\bf n})$ on the faces of the box $\Lambda$.
This contour has also the property of being cut only once by 
all straight lines orthogonal to the diagonal plane 
$i_1+i_2+i_3=0$.
It can therefore be described by specifying
the distance, or height, of each vertex of the unit squares, 
which form the contour, to the plane $i_1+i_2+i_3=0$.
Now, the projection of the cubic lattice ${\bf Z}^3$ onto the 
plane $i_1+i_2+i_3=0$ is a triangular plane lattice, 
which will be denoted by $\cal T$. 
This means that, to each point $t\in{\cal T}$, 
we associate an integer variable $\phi(t)$ which,
multiplied by the factor $1/\sqrt{3}$,
gives the height of the microscopic interface at this point.
The height differences
\beq
n(t,t')=\phi(t)-\phi(t')
\label{A42}\eeq
between nearest neighbors $t$ and $t'$ 
obtained in this way are subject to the following restriction:
Going clockwise along the edges of each elementary triangle 
pointing to the right the values of $n(t,t')$ are $\{1,1,-2\}$. 
An equivalent condition is obtained by using the triangles
pointing to the left and to follow a counterclockwise path
(see Figure 2). 
Conversely, from each configuration $\phi(t)$, $t\in{\cal T}$, 
subject to these restrictions, one can obtain a configuration 
of the microscopic interface.
and, hence, a ground configuration
in ${\cal G}^{(\pm,{\bf n})}(\Lambda)$, 


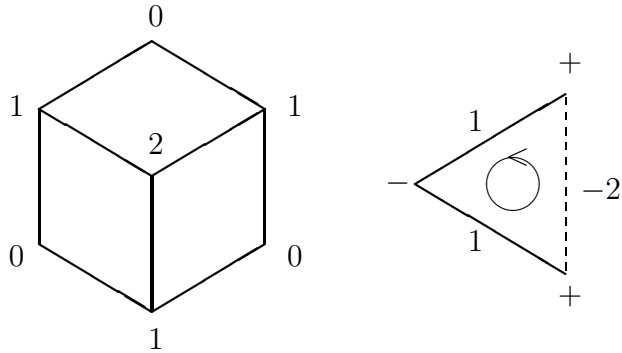
\bigskip\bigskip

\begin{center}
\begin{picture}(70,50)

\thicklines
\put(0,27){\line(5,3){15}}\put(0,27){\line(5,-3){15}}
\put(15,18){\line(5,3){15}}\put(15,36){\line(5,-3){15}}
\put(0,9){\line(5,-3){15}}\put(15,0){\line(5,3){15}}
\put(0,9){\line(0,1){18}}\put(15,0){\line(0,1){18}}
\put(30,9){\line(0,1){18}}
\put(-4,26){$1$}\put(33,26){$1$}
\put(-4,6){$0$}\put(33,6){$0$}
\put(14.5,38){$0$}\put(14.5,21){$2$}\put(14.5,-5){$1$}

\thicklines\put(50,17){\line(5,3){20}}
\put(50,17){\line(5,-3){20}}
\thinlines\multiput(70,5.5)(0,2){12}{\line(0,1){1}}
\put(62,19.5){$<$}\put(63,17){\circle{7}}
\put(72,15){$-2$}\put(57,24){$1$}\put(57,8){$1$}
\put(46,16){$-$}\put(69,32){$+$}\put(69,1){$+$}

\end{picture}
\end{center}

\bigskip\bigskip

\begin{quote}\begin{quote}

{\eightrm Figure 2. (a) Projection of
a cube and heights of the vertices. 
(b) TISOS height differences
as obtained from a ground configuration of the
triangular Ising antiferromagnet.
The corresponding diamond configuration is obtained 
by erasing the edges connecting parallel spins.}

\end{quote}\end{quote}

\bigskip


\bigskip\bigskip

\begin{center}
\begin{picture}(80,60)

\thicklines
\put(-8,-3){\line(0,1){66}}
\put(88,-3){\line(0,1){66}}
\put(-8,-3){\line(1,0){96}}
\put(-8,63){\line(1,0){96}}

\def\aa{\line(5,-3){10}}
\def\bb{\line(5,3){10}}
\def\cc{\line(0,1){12}}

\thinlines
\put(-5,-3){\line(5,3){5}}
\put(-8,4.8){\line(5,-3){8}}
\put(-8,16.8){\line(5,-3){8}}
\put(-8,28.8){\line(5,-3){8}}
\put(-8,31.2){\line(5,3){8}}
\put(-8,4.8){\line(5,-3){8}}
\put(-8,43.2){\line(5,3){8}}
\put(-8,52.8){\line(5,-3){8}}
\put(-8,55.2){\line(5,3){8}}
\put(0,0){\line(5,-3){5}}

\put(0,0){\bb}\put(0,0){\cc}\put(0,12){\bb}\put(0,12){\cc}
\put(0,24){\bb}\put(0,36){\aa}\put(0,36){\bb}\put(0,36){\cc}
\put(0,48){\bb}\put(0,60){\aa}\put(0,60){\line(0,1){3}}

\put(10,6){\cc}\put(10,6){\aa}\put(10,18){\aa}\put(10,18){\cc}
\put(10,30){\aa}\put(10,30){\bb}\put(10,42){\cc}\put(10,42){\aa}
\put(10,54){\aa}\put(10,54){\bb}\put(10,54){\line(0,1){9}}

\put(20,0){\bb}\put(20,0){\cc}\put(20,12){\bb}\put(20,12){\cc}
\put(20,24){\bb}\put(20,36){\aa}\put(20,36){\bb}\put(20,36){\cc}
\put(20,48){\bb}\put(20,60){\aa}\put(20,60){\line(0,1){3}}

\put(30,6){\aa}\put(30,6){\cc}\put(30,18){\cc}\put(30,18){\aa}
\put(30,18){\bb}\put(30,30){\bb}\put(30,42){\cc}\put(30,42){\aa}
\put(30,54){\line(0,1){9}}\put(30,54){\aa}\put(30,54){\bb}

\put(40,0){\bb}\put(40,0){\cc}\put(40,12){\bb}\put(40,24){\cc}
\put(40,24){\aa}\put(40,36){\cc}\put(40,36){\aa}\put(40,48){\aa}
\put(40,48){\bb}\put(40,60){\aa}\put(40,60){\line(0,1){3}}

\put(50,6){\aa}\put(50,6){\cc}\put(50,18){\aa}\put(50,18){\cc}
\put(50,30){\aa}\put(50,30){\bb}\put(50,30){\cc}
\put(50,42){\bb}\put(50,54){\aa}\put(50,54){\line(0,1){9}}
\put(55,63){\line(5,-3){5}}

\put(60,0){\cc}\put(60,12){\aa}\put(60,12){\bb}\put(60,12){\cc}
\put(60,24){\bb}\put(60,36){\aa}\put(60,36){\bb}\put(60,36){\cc}
\put(60,48){\bb}\put(60,48){\cc}
\put(60,60){\line(5,3){5}}

\put(70,6){\bb}\put(70,18){\aa}\put(70,18){\bb}\put(70,18){\cc}
\put(70,30){\bb}\put(70,42){\aa}\put(70,42){\cc}
\put(70,54){\aa}\put(70,54){\line(0,1){9}}
\put(75,63){\line(5,-3){5}}

\put(80,0){\line(5,-3){5}}\put(80,0){\line(5,3){8}}\put(80,0){\cc}
\put(80,12){\line(5,3){8}}
\put(80,24){\line(5,-3){8}}\put(80,24){\cc}
\put(80,36){\line(5,-3){8}}\put(80,36){\line(5,3){8}}\put(80,36){\cc}
\put(80,48){\line(5,3){8}}\put(80,48){\cc}
\put(80,60){\line(5,3){5}}

\put(15,-3){\line(5,3){5}}\put(20,0){\line(5,-3){5}}
\put(35,-3){\line(5,3){5}}\put(40,0){\line(5,-3){5}}
\put(55,-3){\line(5,3){5}}\put(60,0){\line(5,-3){5}}
\put(75,-3){\line(5,3){5}}\put(70,-3){\line(0,1){9}}

\end{picture}
\end{center}

\bigskip

\begin{quote}\begin{quote}

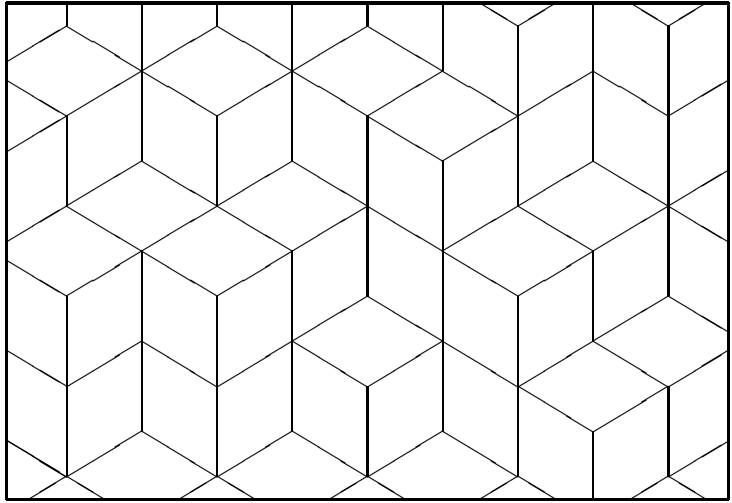
{\eightrm Figure 3. Illustration of
the mapping of a ground microscopic interface 
in 3 dimensions onto a diamond configuration.
A diamond configuration is obtained by erasing one edge
in every elementary triangle of the triangular lattice.}

\end{quote}\end{quote}

\bigskip

\newpage

\noindent
provided that
the boundary conditions are satisfied.
The model with such a set of configurations was introduced 
in ref.\ \cite{BH} and considered also in ref.\ \cite{NHB}.
It is called the {\eightrm TISOS} (triangular Ising 
solid-on-solid) model.

There is a one-to-one correspondence between the 
{\eightrm TISOS} configurations and the  
allowed configurations of the Ising antiferromagnet at zero
temperature.
One can obtain it by requiring that odd and even height 
variables correspond to spins of different sign. 
It is easily checked that this Ising configuration obeys 
the $T=0$ antiferromagnetic constraint that no elementary triangle 
should contain three spins with equal signs.

If one erases all lattice edges between parallel spins, 
one obtains a plane filled with rhombi or diamonds, 
as shown in Figure 3.
Such a diamond covering is just the orthogonal projection 
on the diagonal plane of the edges of the microscopic interface 
initially considered on the cubic lattice.
It gives also a view in perspective of this interface
in the three dimensional space. 

Let ${\bf e}_1$, ${\bf e}_2$ and ${\bf e}_3$ be three vectors
on the plane at angles $2\pi/3$ and modulus equal to $\sqrt{3}/2$.
Since the length of the edges of $\cal T$ is $\sqrt{3}/2$, 
the sites of the lattice $\cal T$ are of the form 
$t_1{\bf e}_1+t_2{\bf e}_2$, with integer coordinates $t_1,t_2$.
Let $Q$ be the set of lattice sites inside the parallelogram 
defined by 
$\vert t_1\vert\le L_1$ and $\vert t_2\vert\le L_2$. 
Let $\vert Q\vert$ be the area of this parallelogram 
and let $\partial Q$ be the boundary of $Q$ (i.e., the set of 
sites in $Q$ having some neighbor outside $Q$).

The boundary conditions corresponding to the slope $u=(u_1,u_2)$
are
\beq
\phi(t)={\bar\phi}_u(t)=[u_1i_1+u_2i_2],\quad i\in\partial Q
\label{A43}\eeq
where $[\cdot]$ denotes the integer part. 
The corresponding projected surface tension is given by 
\beq
\tau_p(u_1,u_2)=\lim_{L_1,L_2\to\infty}-{1\over{\beta\vert Q\vert}}
Z_\Lambda(u_1,u_2)
\label{A44}\eeq
with
\beq
Z_\Lambda(u_1,u_2)=\sum e^{-\beta H(\phi)}
\label{A45}\eeq
In (\ref{A45}) the sum runs over all configurations $\phi$ 
inside $Q$
satisfying the boundary conditions (\ref{A43}), and $H(\phi)= 
H_{\Lambda}(\sigma_\phi\vert(\pm,{\bf n}))-H_{\Lambda}(+\vert(+))$
is the relative energy of the ground configuration $\sigma_\phi$ 
associated to the microscopic interface defined by $\phi$.

We can interpret the conditions (\ref{A43}) as ``canonical'' 
constraints 
and introduce a conjugate Gibbs ensemble of (\ref{A45}), which can
be viewed as a ``grand canonical'' ensemble with respect to 
the interface boundaries. 
For this purpose we consider the boundary terms
\beq
S_1(\phi) =\sum_{\ell\in\ell_1(Q)}n(\ell)\ ,\quad
S_2(\phi) =\sum_{\ell\in\ell_2(Q)}n(\ell) 
\label{A46}\eeq
where $\ell_1(Q)$ and $\ell_2(Q)$ are the sets of all bonds 
in $Q$ parallel to the vectors ${\bf e}_1$ and ${\bf e}_2$, 
respectively, and oriented according to increasing coordinates.
The bond variables $n(\ell)$ are the height differences of
formula (\ref{A42}). 
The grand canonical prescription, which is convenient to consider,
consists in adding to the energy a term of the form
\beq
x_1S_1(\phi)+x_2S_2(\phi)
\label{A47}\eeq
where $x=(x_1,x_2)\in{\bf R}^2$ represent the slope ``chemical'' 
potentials.
The associated partition function and free energy are 
\begin{eqnarray}
\Xi^{\rm per}_Q(x_1,x_2) &=&\sum_{\phi}e^{-\beta H(\phi)+
\beta x_1S_1(\phi)+\beta x_2S_2(\phi)} \label{A48}\\
\varphi^{\rm per}_Q(x_1,x_2) &=&
-{1\over{\beta\vert Q\vert}}\ln\, 
\Xi^{\rm per}_Q(x_1,x_2) \label{A49}
\end{eqnarray}
The sum in (\ref{A48}) runs over all configurations in $Q$ with 
periodic boundary conditions (with respect to the bond
variables $n(\ell)$), $\phi(0)$ is taken equal to 0.

\medskip

{\bf Theorem 2. }{\it
The limit, when $Q\to\infty$, of 
$\varphi^{\rm per}_Q$, exists and equals
\beq
\varphi(x_1,x_2)=-\sup_u\{x_1u_1+x_2u_2-\tau_p(u_1,u_2) \}
\label{A50}\eeq
in the interior of the domain of the concave function $\varphi$.}

\medskip

The theorem is proved in ref.\ \cite{MR} 
for many {\eightrm SOS} models.
Notice that the partition function (\ref{A49}) does not correspond 
exactly to the conjugate ensemble of the function (\ref{A45}),  
as conditions (\ref{A43}) represent many more constraints that 
just fixing the values of the two sums in (\ref{A46}).
The domain of a concave function $\varphi(x)$ is the set
$\{x\in{\bf R}^2 : \varphi(x)>-\infty\}$.

It turns out that the free energy function $\varphi(x)$
of the {\eightrm TISOS} model can be computed exactly. 
There is no additional difficulty in considering, 
like in section 1, an asymmetric Ising model 
on the cubic lattice (with three different interaction parameters 
in the vertical and the two horizontal directions). 
However, it is interesting instead 
to introduce a new kind of asymmetry, 
and to take the interaction parameters equal to 
$J_1$, $J_2$ or $J_3$,
according to whether the coordinates of the highest point
of the corresponding bond satisfy $i_1+i_2+i_3\equiv 
0,1,\hbox{ or }2$ (mod 3).

Then, with $K_i=\beta J_i$ for $i=1,2,3$, we write
\begin{eqnarray}
r_1&=&\exp\,(-K_1+K_2+K_3)\nonumber\\
r_2&=&\exp\,(K_1-K_2+K_3)\nonumber\\  
r_3&=&\exp\,(K_1+K_2-K_3)\nonumber\\
r_0^3&=&r_1r_2r_3 \ =\ \exp\,(K_1+K_2+K_3) \label{A51}
\end{eqnarray}
and
\beq
R=(r_1^3+r_2^3+r_3^3)r_0^{-3}
\label{A52}\eeq

For reasons of symmetry, the grand canonical prescription, 
which will be considered,
consists in adding to the energy, instead of the term (\ref{A47}), 
a term of the form
\beq
x_1S_1(\phi)+x_2S_2(\phi)+x_3S_3(\phi),\quad 
\hbox{with } S_3(\phi) =\sum_{\ell\in\ell_3(Q)}n(\ell) 
\eeq
where $x_3\in{\bf R}$  
and $\ell_3(Q)$ is the set of all bonds in $Q$ parallel
to the vector ${\bf e}_3$.
Taking, $X_i=\beta x_i$, for $i=1,2,3$, we write
\begin{eqnarray}
w_1&=&\exp\,(-X_1+X_2+X_3)\nonumber\\
w_2&=&\exp\,(X_1-X_2+X_3)\nonumber\\
w_3&=&\exp\,(X_1+X_2-X_3)\nonumber\\
w_0^3&=&w_1w_2w_3\ =\ \exp\,(X_1+X_2+X_3) \label{A53}
\end{eqnarray}

Then, we have
\begin{eqnarray}
&&\varphi\ =\ \lim_{L_1,L_2\to\infty}{2\over{L_1L_2}}\ln\,\Xi 
\ =\ \ln (w_0/r_0) +\nonumber\\
&&{1\over{24\pi^2}}\int_0^{2\pi}dp\int_0^{2\pi}dq\,
\ln\,\big\vert R w_0^3 e^{ip} 
+w_1^3e^{3ip}-w_2^3e^{-3iq}+w_3^3e^{3iq}\big\vert^2 \label{A55}
\end{eqnarray}

The exact solution (\ref{A55}) for the free energy of the model
is due to Nienhuis et al.  \cite{NHB} 
(where slightly different notations are used).
A careful study of this solution, also in ref.\ \cite{NHB},
shows the existence of four regions in which the function
$\varphi$ has simple expressions, as indicated below
$$\matrix{
\hbox{1)\quad if}
&\hfill -w_1^3+w_2^3+w_3^3+R\,w_0^3\le0,
&\varphi=\ln w_1 +\ln\,(w_0/r_0) \kern 1cm
\hfill\cr\cr
\hbox{2)\quad if}
&\hfill w_1^2-w_2^3+w_3^3+R\,w_0^3\le0,
&\varphi=\ln w_2 + \ln\,(w_0/r_0) \hfill\cr\cr
\hbox{3)\quad if}
&\hfill w_1^3+w_2^3-w_3^3+R\,w_0^3\le0,
&\varphi=\ln w_3 + \ln\,(w_0/r_0) \hfill\cr\cr
\hbox{4)\quad if}
&\hfill w_1^3+w_2^3+w_3^3-R\,w_0^3\le0,
&\varphi=\varphi_0 \hfill\cr
}$$

\medskip

In regions 1), 2) and 3), $\varphi$ is a linear function 
of the variables $x_1,x_2,x_3$. 
It is constant in region 4), where it takes the value of
(\ref{A55}) at the point $x_1=x_2=x_3=0$. 
These regions correspond to the domains of four ordered
phases of the {\eightrm TISOS} model in the phase diagram
of this system.

According to the Andreev construction, 
the graph of the function
\beq
z=-(1/\beta)\,\varphi(x_1,x_2) 
\label{A56}\eeq
gives the boundary of the Wulff shape for the equilibrium
crystal (equation (\ref{A10})). 
The factor $-1/\beta$ comes from definition (\ref{A55})
in order to obtain the correct free energy.
We have, in region 3), for instance,
\beq
z=J_1+J_2+J_3-2(x_1+x_2)
\label{A57}\eeq
which shows that the Wulff shape contains a plane 
facet orthogonal to the (0,0,1) direction. 
The boundary of this facet is determined by 
the boundary of region 3). 
The other regions 1), 2) and 4) indicate the existence 
of three other facets respectively orthogonal to the 
(1,0,0), (0,1,0) and (1,1,1) directions. 
The portion outside these regions corresponds to the curved 
part of the crystal surface. 

It is only in recent times that equilibrium crystals 
have been produced in the laboratory,  
information and references on this subject can be found 
in the review articles \cite{rottman} and \cite{bejeren}. 
As in the present model,   
a typical equilibrium crystal at low 
temperatures has smooth plane facets 
linked by rounded edges and corners.
The area of a particular facet decreases as the 
temperature is raised and the facet finally disappears 
at a temperature characteristic of its orientation, 
a phenomenon known under the name of roughening transition.
The reader will find in ref.\ \cite{NHB} a study of 
the roughening transition for the (1,1,1) facet 
in the {\eightrm TISOS} model, as well as a 
discussion on several other interesting properties 
of the crystal shape. 

In the symmetric case, in which the interaction parameters
of the Ising model are equal, $J_1=J_2=J_3=J$, we have $R=3$
(from equations (\ref{A51}) and (\ref{A52})). 
Then the region of phase 4) is empty
and the (1,1,1) facet does not exist. 
This follows from the algebraic identity  
\begin{eqnarray}
&&w_1^3+w_2^3+w_3^3-3w_1w_2w_3 = \nonumber\\
&&(w_1+w_2+w_3)(w_1^2+w_2^2+w_3^2-w_1w_2-w_2w_3-w_3w_1) = \nonumber\\
&&{\scriptstyle{1\over2}}(w_1+w_2+w_3)
((w_1-w_2)^2+(w_2-w_3)^2+(w_3-w_1)^2) \label{A58} 
\end{eqnarray}
showing that
\beq
w_1^3+w_2^3+w_3^3-3w_0^3\ge 0 
\label{A59}\eeq 

Conversely, in the case that $J_1$, $J_2$ and $J_3$ are 
not equal one has $R>3$ and expression (\ref{A59}) 
can be negative.
Then region 4) is not empty if $\beta$ is large enough, 
which implies the appearance of the (1,1,1) facet at 
low temperatures.

\bigskip\bigskip

\begin{center}
\begin{picture}(40,40)

\thinlines
\put(0,40){\line(4,-3){20}}
\put(20,25){\line(4,3){20}}
\put(20,0){\line(0,1){25}}
\thicklines
\bezier{200}(2.25,40)(16.25,29.5)(20,29.5)
\bezier{200}(1,38)(15,26)(16.5,23)
\bezier{200}(16.5,23)(18,20)(18.5,1)
\bezier{200}(20,29.5)(23.75,29.5)(37.75,40)
\bezier{200}(23.5,23)(25,26)(39,38)
\bezier{200}(21.5,1)(22,20)(23.5,23)
\put(19,37){$\bf 3$}\put(9,17){$\bf 1$}\put(29,17){$\bf 2$}
\put(-4,42){$x_2$}\put(19,-3){$x_3$}\put(41,42){$x_1$}

\end{picture}
\end{center}

\bigskip

\begin{quote}\begin{quote}
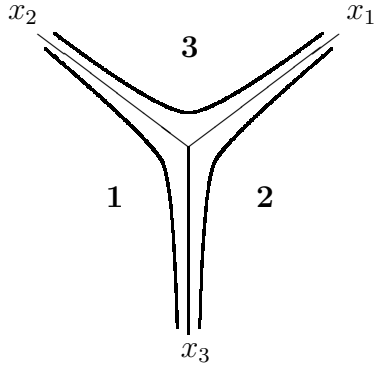
{\eightrm Figure 4. Equilibrium crystal shape according to the
TISOS model in the symmetric case.
The crystal is shown in a projection parallel to 
the (1,1,1) direction.
The three regions 1), 2) and 3) indicate the facets, 
and the remaining area represents the curved part 
of the crystal surface.}

\end{quote}\end{quote}

\bigskip

Figure 4 shows a (1,1,1) corner of the crystal for
the symmetric case $J_1=J_2=J_3=J$. 
In this case the definition of region 3), for instance, 
can be written, using again identity (\ref{A58}), 
\begin{eqnarray}
&&w_1^3+w_2^3-w_3^3+3w_1w_2w_3 = \nonumber\\
&&{\scriptstyle{1\over2}}(w_1+w_2-w_3)
((w_1-w_2)^2+(w_2+w_3)^2+(-w_3-w_1)^2)\le0 \label{A60} 
\end{eqnarray}
or, equivalently,
\beq
w_1+w_2-w_3\le 0 \label{A61}
\eeq
and, using (\ref{A53}) with $x_3=0$, we see that region 3) 
can be defined by 
\beq
e^{-2\beta x_1}+e^{-2\beta x_2}\le 1 \label{A62}
\eeq
The comparison of this formula with equation (\ref{A34}) 
shows the following statement.

\medskip

{\tt Remark. }
The shape of the facets of the equilibrium crystal associated
with the {\eightrm TISOS} model, in the symmetric case, 
coincide with the Wulff shape obtained from the approximate 
solution of the two dimensional Ising model. 

\medskip

This property can be derived, without using the exact 
solution (\ref{A55}), with the help of the results of 
section 4 below, in particular Corollary 1.
These results apply also to the {\eightrm TISOS} model, 
and the shape of the facet is obtained by computing the 
corresponding $\tau^{\rm step}({\bf m})$. 
Notice also that 
the model can be used to describe the (1,1,1) corner of
the crystal, that is 1/8 of the crystal. 
Completing it by symmetry we see that its facets 
undergo a roughening transition at $K'_c=(1/2)\ln2=0.346574$
(see last paragraph of section 2), and 
we have $K'_c>K_c{(3)}$, where $K_c{(3)}\sim0.22$ 
is the value of the critical coupling constant for the 
three dimensional Ising model.

There are other models for which an analysis similar 
to that of the present section could be developed. 
For example, the Ising model on a body centered cubic 
lattice with nearest and next nearest neighbor interactions
is one of these models.
This system leads also to an approximate solution that
can be expressed in terms of exactly solvable 
{\eightrm SOS} models 
(see \cite{KM} and the references quoted there).

\section{The three dimensional model: \hfill\break 
Rigorous results}

In the previous section we have described the microscopic 
interfaces at zero temperature for the Ising model 
on a cubic lattice. 
Some properties of these interfaces were studied by means
of the exact solution of an associated {\eightrm SOS} model. 
At low (positive) temperature we expect the ground interfaces 
to be modified by deformations.
Small deformations will appear here and there, the 
large deformations having a very small probability. 
It appears, however, very difficult to develop such an argument
into a rigorous proof that could justify the validity at low
temperatures of the approximate solution described
in section 3. 

The situation is different for the interface orthogonal 
to the direction ${\bf n}_0=(0,0,1)$. 
In this case there is, at zero temperature, only one 
microscopic interface, which coincides with the horizontal
plane. 
The small deformations of this interface that appear at 
low temperatures are accessible to a mathematical treatment. 
The interface is rigid at low temperatures and its
properties can be studied by means of a convergent 
cluster expansion. 
By pursuing this analysis it has been possible to determine
the shape of the facets in a rigorous way. 
It can then be seen \cite{M}, from the proof of Theorem 3 below 
and Corollary 1, that this shape differs from the 
Wulff shape associated with the two dimensional Ising
model, only by a quantity of order $\exp(-12\beta J)$, 
and thus, that it is close to the shape predicted by 
the approximate solution of section 3.

The appearance of a facet 
in the equilibrium crystal shape is related, 
according to the Wulff construction, to the existence 
of a discontinuity in the derivative of the surface
tension with respect to the orientation. 
More precisely, 
assume that the surface tension satisfies
the convexity condition of Theorem 1, and let 
this function $\tau({\bf n})=\tau(\theta,\phi)$
be expressed in terms of the spherical co-ordinates
$0\le\theta\le\pi$, $0\le\phi\le 2\pi$ of ${\bf n}$, 
the vector ${\bf n}_0$ being taken as the $x_3$ axis.
A facet orthogonal to the direction ${\bf n}_0$ appears
in the Wulff shape ${\cal W}$
if, and only if, the derivative
$\partial\tau(\theta,\phi)/\partial\theta$
is discontinuous at the point $\theta=0$,
for all $\phi$. 
Moreover,
the one-sided derivatives with respect to
$\theta$ exist,
at $\theta=0^+$ and $\theta=0^-$, 
and determine the shape of the facet. 
The facet ${\cal F}\subset\partial{\cal W}$ 
consists of the points ${\bf x}\in{\bf R}^3$
belonging to the plane $x_3=\tau({\bf n}_0)$ and such that
\beq
x_1\cos\phi+x_2\sin\phi\le
\partial\tau(\theta,\phi)/\partial\theta\,\vert\,_{\theta=0^+}  
\label{B1}\eeq
for every $\phi$ between $0$ and $2\pi$. 
A proof of this fact can be found in ref.\ \cite{M} (Theorem 1).

The step free energy is expected to play an important role 
in the facet formation.
It is defined as the free energy (per unit length)
associated with the introduction of a step of height 1   
on the interface,  
and can be regarded as an order parameter
for the roughening transition. 
Let $\Lambda$ be a parallelepiped of sides 
$L_1,L_2,L_3$, parallel to the axes,
and centered at the origin, and introduce the 
$(\hbox{\eightrm step},{\bf m})$
boundary conditions, associated to the unit vectors
${\bf m} = (\cos\phi,\sin\phi)\in{\bf R}^2$, by
\beq
{\bar\sigma}(i) = 
\cases{
1  &if\ \ $i>0$\ \ or if\ \ $i_3=0$\ \ and\ \ $i_1m_1+i_2m_2\ge0$\cr
-1  &otherwise \cr}  
\label{B2}\eeq
Then, the step free energy, for a step
orthogonal to ${\bf m}$ (with $m_2\ne0$) 
on the interface orthogonal to ${\bf n}_0=(0,0,1)$, is   
\beq 
\tau^{\rm step}(\phi) =
\lim_{L_1\to\infty}\lim_{L_2\to\infty}\lim_{L_3\to\infty}
- {{\cos\phi}\over{\beta L_1}}\  
\ln\  {{Z^{({\rm step},{\bf m})}(\Lambda)}\over 
{Z^{(\pm,{\bf n}_0)}(\Lambda)}} 
\label{B3}\eeq

A first result concerning this point,   
was obtained by Bricmont et al.\ \cite{BMF}.  
These authors proved
a correlation inequality which establish
$\tau^{\rm step}(0)$ 
as a lower bound to the one-sided derivative
$\partial\tau(\theta,0)/\partial\theta$ at $\theta=0^+$
(the inequality extends also to $\phi\ne0$, see \cite{M}).     
Thus when $\tau^{\rm step}>0$ a facet is expected. 

Using the perturbation theory of the horizontal interface,  
it is possible, as mentioned above, to study also the 
microscopic interfaces associated with the 
$(\hbox{\eightrm step},{\bf m})$ boundary conditions. 
The step structure at low temperatures can then be 
analyzed with the help of a new cluster expansion. 
As a consequence of this analysis we have 
the following theorem.

\medskip

{\bf Theorem 3. } {\it 
If the temperature is low enough,  
i.e., if $\beta\ge\beta_0$, where $\beta_0>0$ is a given constant, 
then the step free energy $\tau^{\rm step}(\phi)$,
defined by limit (\ref{B3}), exists, is strictly positive, and 
extends by positive homogeneity to a strictly convex function.
Moreover, $\tau^{\rm step}(\phi)$ can be expressed as an
analytic function of $\beta$, 
for which a convergent series expansion can be found.}

\medskip

Using the above results on the step structure, 
similar methods allow us to evaluate
the increment in surface tension of an interface 
titled by a very small angle $\theta$ 
with respect to the rigid horizontal interface. 
This increment can be expressed in terms of the 
step free energy and one obtains the following relation. 
                                           
\medskip

{\bf Theorem 4. } {\it
For $\beta\ge\beta_0$, we have}
\beq
\partial\tau(\theta,\phi)/\partial\theta\,\vert\,_{\theta=0^+}
= \tau^{\rm step}(\phi)   
\label{B4}\eeq

\medskip

This relation, together with equation (\ref{B1}),
implies that one obtains the shape of the facet 
by means of the two-dimensional Wulff construction
applied to the step free energy.

\medskip

{\bf Corollary 1. }{ \it The equilibrium crystal presents, 
if $\beta\ge\beta_0$, a facet orthogonal to the (0,0,1) 
direction, whose shape is given by }
\beq
{\cal F} =\left\{{\bf x}\in{\bf R}^2 : 
{\bf x}\cdot{\bf m}\le\tau^{\rm step}({\bf m})  
\hbox{\ \ for every\ \ }{\bf m}\in{\bf S}^1\right\}
\label{B5}\eeq

\medskip

The reader will find a detailed discussion on these points, 
as well as the proofs of Theorems 3 and 4, in ref.\ \cite{M}. 
A brief explanation of these results can be found in 
ref.\ \cite{Mb}.

From the properties of $\tau^{\rm step}$ stated in Theorem 3, 
it follows that the Wulff equilibrium crystal 
presents well defined boundary lines, smooth and  
without straight segments, between
a rounded part of the crystal surface and the facets parallel
to the three main lattice planes.

It is expected, but not proved, that at a higher temperature, 
but before reaching the critical temperature, 
the facets associated with the Ising model undergo a 
roughening transition. The conjectured value of
$\beta_R$, the roughening inverse temperature, 
is given by $K_R\sim0.38$ \cite{WGM}.
It is then natural to believe that the equality (\ref{B4}) 
is true for any $\beta$ larger than $\beta_R$,   
allowing us to determine the facet shape from  
Corollary 1, 
and that for $\beta\le\beta_R$,  
both sides in the equality vanish,
and thus, the disappearance of the facet is involved.
However, the condition that the temperature 
is low enough is needed in the proofs of the above results.

\end{document}